\documentclass[twocolumn,aps,prl,showpacs,amsmath,amssymb]{revtex4}

\usepackage{graphicx}
\usepackage{dcolumn}
\usepackage{bm}

\begin{document}

\title{Anisotropic Magnetoresistance in Lightly Doped
La$_{2-x}$Sr$_x$CuO$_4$: Impact of Anti-Phase Domain Boundaries on the
Electron Transport}

\author{Yoichi Ando}
\author{A. N. Lavrov}
\author{Seiki Komiya}

\affiliation{Central Research Institute of Electric Power Industry,
Komae, Tokyo 201-8511, Japan}

\date{\today}

\begin{abstract}

Detailed behavior of the magnetoresistance (MR) is studied in lightly
doped antiferromagnetic La$_{1.99}$Sr$_{0.01}$CuO$_4$, where,
thanks to the weak ferromagnetic moment due to spin canting, the
antiferromagnetic (AF) domain structure can be manipulated by the magnetic
field. The MR behavior demonstrates that CuO$_2$ planes indeed contain
anti-phase AF domain boundaries in which charges are confined, forming
anti-phase stripes. The data suggest that a high magnetic field turns
the anti-phase stripes into in-phase stripes, and the latter appear to
give better conduction than the former, which challenges the notion that
the anti-phase character of stripes facilitates charge motion.

\end{abstract}


\maketitle

In high-$T_c$ cuprates, there is growing evidence that charges and spins
self-organize in CuO$_2$ planes in a peculiar striped manner, where the
doped holes are arranged in fluctuating lines, ``charged stripes'', that
separate antiferromagnetic (AF) domains \cite{theory,NdSr,Yamada}.
This intriguing microscopic state has been proposed to be responsible
for many unusual properties of cuprates
\cite{SC,mobility,Hall,suscept,anis,ourstr}, but information
on the role of stripes is still quite scarce.

Recently, it has been found that lightly doped YBa$_2$Cu$_3$O$_{6+x}$
(YBCO) and La$_{2-x}$Sr$_{x}$CuO$_4$ (LSCO) crystals develop a remarkable
{\it in-plane} resistivity anisotropy upon decreasing temperature, which
has been attributed to the self-organization of holes into unidirectional
conducting stripes \cite{anis}. Moreover, in YBCO, an application of the
magnetic field induces a persistent change in the in-plane anisotropy,
presumably caused by some rearrangement of the stripes \cite{ourstr}. If
this field-induced phenomenon is indeed related to the inherent striped
structure of CuO$_2$ planes \cite{discus}, similar features should be
generic to cuprates, and by manipulating the stripes with a magnetic field
one should be able to gain insights into their roles in macroscopic
properties.

It is thus natural to turn to the LSCO system, where clear
unidirectional striped structure has been observed by neutron scattering
\cite{1D}. What makes LSCO even more attractive for the
magnetoresistance study is a weak ferromagnetic (FM) component that
always accompanies the AF order: In LSCO, the spins in CuO$_2$ planes
are slightly canted from the direction of the staggered magnetization,
providing a weak FM moment whose direction is uniquely linked with the
{\it phase} of the AF order \cite{suscept,MR_WF,MR_flop}. As a result,
once CuO$_2$ planes develop a pattern of AF domains that are separated
by anti-phase boundaries \cite{NdSr,suscept,1D}, the same pattern of FM
moments emerges as well. Apparently, by using an external magnetic field
one should be able to manipulate the domain structure in LSCO in quite
the same way as in usual ferromagnets \cite{FM}; a large enough field
should drive undoped or lightly-doped LSCO into a weak-ferromagnetic
state \cite{MR_WF}, where all the weak FM moments are aligned and the
magnetic domain boundaries, if any, are completely wiped out. The
consequent resistivity evolution would indicate how the AF-domain
boundaries (stripes) are important for the electron transport.

Following this anticipation, we study the anisotropic magnetoresistance
(MR) in lightly doped, antiferromagnetic La$_{1.99}$Sr$_{0.01}$CuO$_4$
single crystals, and find that a large magnetic field actually has a
significant influence on the charge transport. The change in the
in-plane and out-of-plane resistivity ($\rho_{ab}$ and $\rho_c$) at 14 T
is observed to be as large as a factor of two and four, respectively, at
low temperature. In particular, the in-plane MR behavior demonstrates
that in zero field each CuO$_2$ plane indeed contain anti-phase domain
boundaries, and that the high magnetic field unifies the phase of the AF
ordering and wipes out the phase boundaries. We argue that the holes in
the phase-unified state in high magnetic fields are still confined in
stripes, which necessarily constitute in-phase domain boundaries. Thus,
in lightly-doped LSCO, the magnetic field has an intriguing function of
switching the stripes from anti-phase boundaries to in-phase ones.

The high-quality La$_{2-x}$Sr$_x$CuO$_4$ single crystals are grown by the
traveling-solvent floating-zone technique and carefully annealed in pure
helium to remove excess oxygen. Resistivity measurements are carried out
by the ac four-probe method on samples that are cut and polished into
suitable shapes. The particular samples reported here were made to be a
thin strip $3000\times 420 \times 70$ $\mu$m$^3$ for $\rho_{ab}$ and a
narrow bar $115\times 280\times 2200$ $\mu$m$^3$ for $\rho_c$. Upon
measuring $\rho_{ab}$, a special care is paid to avoid an admixture of
$\rho_c$: upon thinning the sample, its face is adjusted to the $ab$
crystal plane with an accuracy better than 1$^{\circ}$, and the current
contacts are carefully placed to cover the sample's side faces. Also,
since we have recently found that the crystallographic twins in
high-quality LSCO crystals can move under applied magnetic fields
\cite{Nature} and it is hard to control this effect, we cut the sample at
45$^\circ$ to the orthorhombic $a$ and $b$ axes ({\it i.e.}, along the
Cu-O-Cu direction) so that the $\rho_a$ and $\rho_b$ components are
averaged. The MR is measured by sweeping the magnetic field at fixed
temperatures stabilized by a capacitance sensor with an accuracy of $\sim
1$ mK. The angular dependence of the MR is determined by rotating the
sample within a 200$^{\circ}$ range under constant magnetic fields of
$\pm14$ T. Magnetization measurements are performed on large ($\sim 0.5$
g) detwinned single crystals \cite{suscept}.

\begin{figure}[t]
\includegraphics[clip,width=6.1cm]{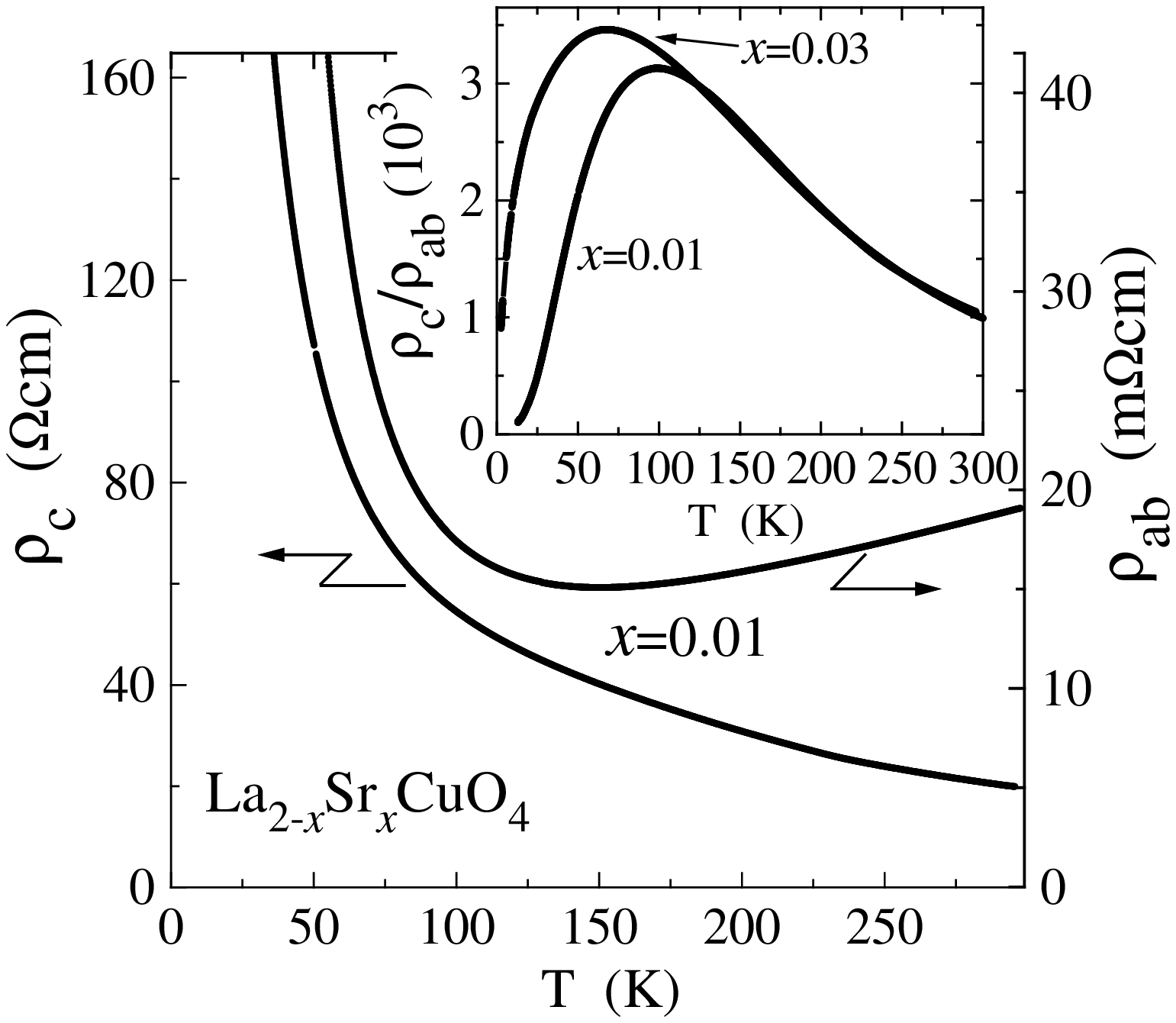}
\caption{$\rho_{c}(T)$ and $\rho_{ab}(T)$ of La$_{2-x}$Sr$_x$CuO$_4$
($x=0.01$) single crystals used for the MR measurements.
Inset: Resistivity anisotropy
$\rho_{c}/\rho_{ab}$ for LSCO crystals with $x=$ 0.01 and 0.03.}
\label{fig1}
\end{figure}

Although doping 1\% of holes into CuO$_2$ planes is not enough to suppress
the AF order ($T_N$ is still 230-240 K \cite{mobility}), it nevertheless
results in appearance of a metal-like in-plane conduction at moderate
temperatures (Fig. 1). However, the mechanism that facilitates the charge
motion within CuO$_2$ planes is apparently irrelevant to the out-of-plane
transport, causing the transport to be quasi-2D with the resistivity
anisotropy $\rho_{c}/\rho_{ab}$ of up to several thousands (inset of Fig.
1). Note that the 2D-conductivity features are essentially a
finite-temperature property at this low doping, since the anisotropy
sharply diminishes as $\rho_{ab}(T)$ loses its metal-like behavior at low
temperatures (for $x=0.01$, $\rho_{c}/\rho_{ab}\approx100$ at $T=13$ K).

\begin{figure}[!t]
\includegraphics[clip,width=8.6cm]{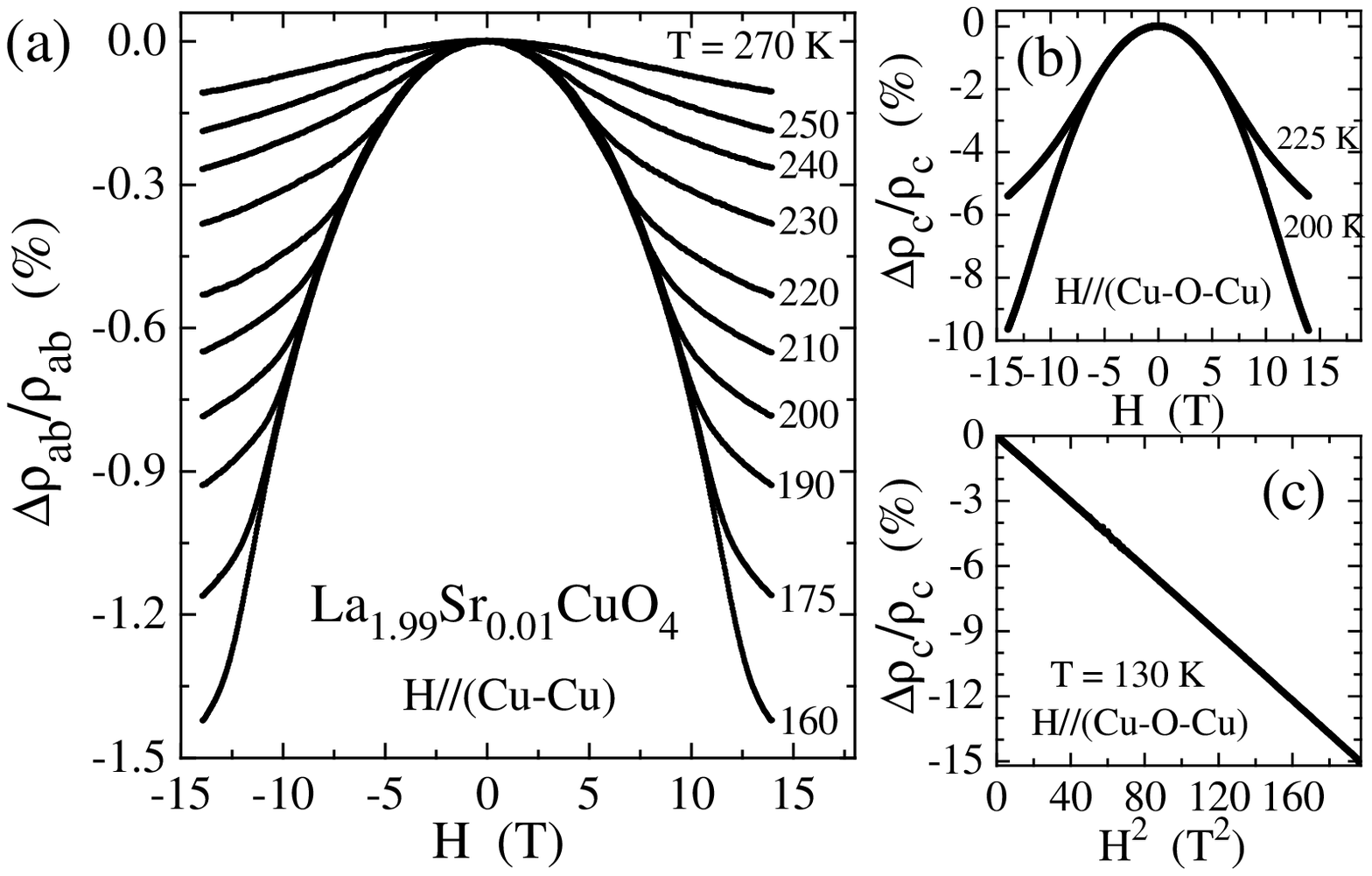}
\caption{(a) The MR in $\rho_{ab}$ of La$_{1.99}$Sr$_{0.01}$CuO$_4$ for the
in-plane magnetic field applied along the Cu-Cu (diagonal) direction,
which is 45$^\circ$ to the current ($I\parallel$ Cu-O-Cu).
(b) The MR in $\rho_{c}$ for $H\parallel$ Cu-O-Cu at high temperatures.
(c) $\Delta\rho_{c}/\rho_{c}$ at 130 K as a function of $H^2$.}
\label{fig2}
\end{figure}

The in-plane MR, $\Delta\rho_{ab}/\rho_{ab}$, measured in
La$_{1.99}$Sr$_{0.01}$CuO$_4$ crystals for the magnetic field $H$
applied parallel to CuO$_2$ planes [Fig. 2(a)] surprisingly resembles that
of lightly doped YBCO \cite{ourstr}; namely, the MR is negative, follows
a $T$-independent curve at low fields, and tends to saturate above some
threshold field $H_{th}$. This similarity, being present despite a
notable distinction between LSCO and YBCO in both the crystal and
magnetic structures \cite{suscept,MR_WF}, demonstrates that the observed
MR behavior is inherent in the lightly doped CuO$_2$ planes.
Nevertheless, there are also several important differences between the
MR features in LSCO and YBCO. First, in LSCO the saturation field as
well as the MR values are scaled up so that
$\Delta\rho_{ab}/\rho_{ab}$ at 14 T exceeds $1\%$ already at high
temperatures and reaches $30-40\%$ at 10 K (in comparison with
$\sim 1\%$ in YBCO \cite{ourstr}); thus, the impact of the magnetic
field is no more weak. Second, the angular dependences of the MR look
quite different: While in YBCO $\Delta\rho_{ab}/\rho_{ab}$ changes its
sign in a $d$-wave manner upon rotating the magnetic field within the
$ab$ plane \cite{ourstr}, in LSCO it is always negative.

Detailed angular dependence study reveals that both the in-plane and
out-of-plane MR of LSCO exhibit a clear two-fold symmetry upon rotating
the magnetic field within the $ab$ plane (Figs. 3 and 4), which is
particularly evident for the $\rho_{c}$-crystal that is almost
single-domain according to x-ray data. The $\Delta\rho_{c}/\rho_{c}$ data
for 130 K shown in Fig. 3 are surprisingly well described by a simple
$A+B\sin^2\alpha$ dependence. Apparently, this simple $\sin^2\alpha$
dependence, combined with the perfect $\Delta\rho_{c}/\rho_{c} \propto
H^2$ behavior observed at 130 K [Fig. 2(c)], indicate that {\it only the
magnetic field component along the orthorhombic $b$-axis, $H_b=H\sin
\alpha$, is responsible for the angular-dependent part of the MR}. The MR
follows the $\sin^2\alpha$ curve as long as the magnetic field stays below
the saturation field $H_{th}$; at 210 K, $H_{th}$ is reduced to $9-10$ T
[Fig. 2(b)], and the MR measured at $H=14$ T shows a constant value over
some range of angles (Fig. 3).

\begin{figure}[!t]
\includegraphics[clip,width=6.0cm]{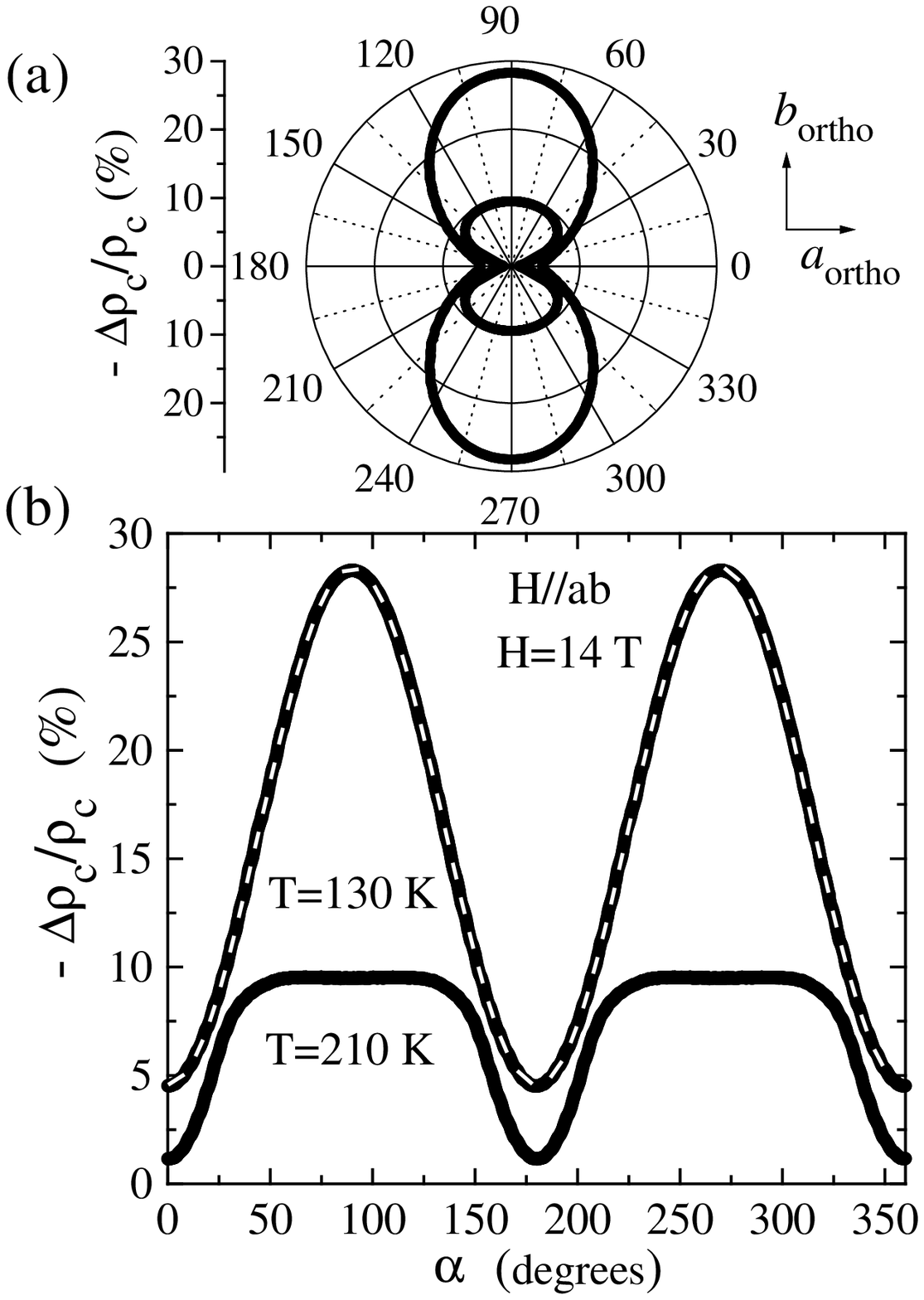}
\caption{Angular dependences of $\Delta\rho_c/\rho_c$ at $T=130$ K and
210 K in (a) polar and (b) linear coordinates (arrows indicate the
directions of the orthorhombic crystal axes). The white dashed line in (b)
shows the fit with $\Delta\rho_c/\rho_c=A+B\sin^2\alpha$.}
\label{fig3}
\end{figure}

The behavior of $\Delta\rho_{ab}/\rho_{ab}$ in Fig. 4 looks different from
that of $\Delta\rho_c/\rho_c$, partly because there are two types of
orthogonal crystallographic domains (twins). Also, the $H$ dependence of
$\Delta\rho_{ab}/\rho_{ab}$ turns out to be $\propto H^n$ with $n>2$ for
$H < H_{th}$ \cite{note}, and this ``anharmonicity'' is responsible for
the rather complicated angular dependence. However, by using the
experimental data of $\Delta\rho_{ab}/\rho_{ab}(H)$ at $H\parallel b$
[Fig. 2(a)] and the ratio of crystallographic domains, we can well
reproduce the observed angular dependence of the MR at both 130 K and 210
K in Fig. 4, assuming that only the $b$-component of magnetic field is
responsible. Thus, in both $\rho_{ab}$ and $\rho_c$ the angular dependence
of the MR is governed solely by the $b$-component of $H$ when the field is
applied in-plane.

When the magnetic field is applied along the $c$-axis, both the in-plane
and out-of-plane MR exhibit a step-like decrease [Fig. 5(a)] similar to
that reported by Thio {\it et al.} for La$_2$CuO$_{4+\delta}$
\cite{MR_WF}. The salient point here is the magnitude of the MR:
$|\Delta\rho_{ab}/\rho_{ab}|$ grows up to $\sim50\%$ at low temperatures
[Fig. 5(b)] and $|\Delta\rho_c/\rho_c|$ reaches $\sim75\%$ [Fig. 5(c)],
that is, the magnetic field is capable of reducing the resistivity by a
factor of 2 and 4, respectively.

\begin{figure}[!t]
\includegraphics[clip,width=6.0cm]{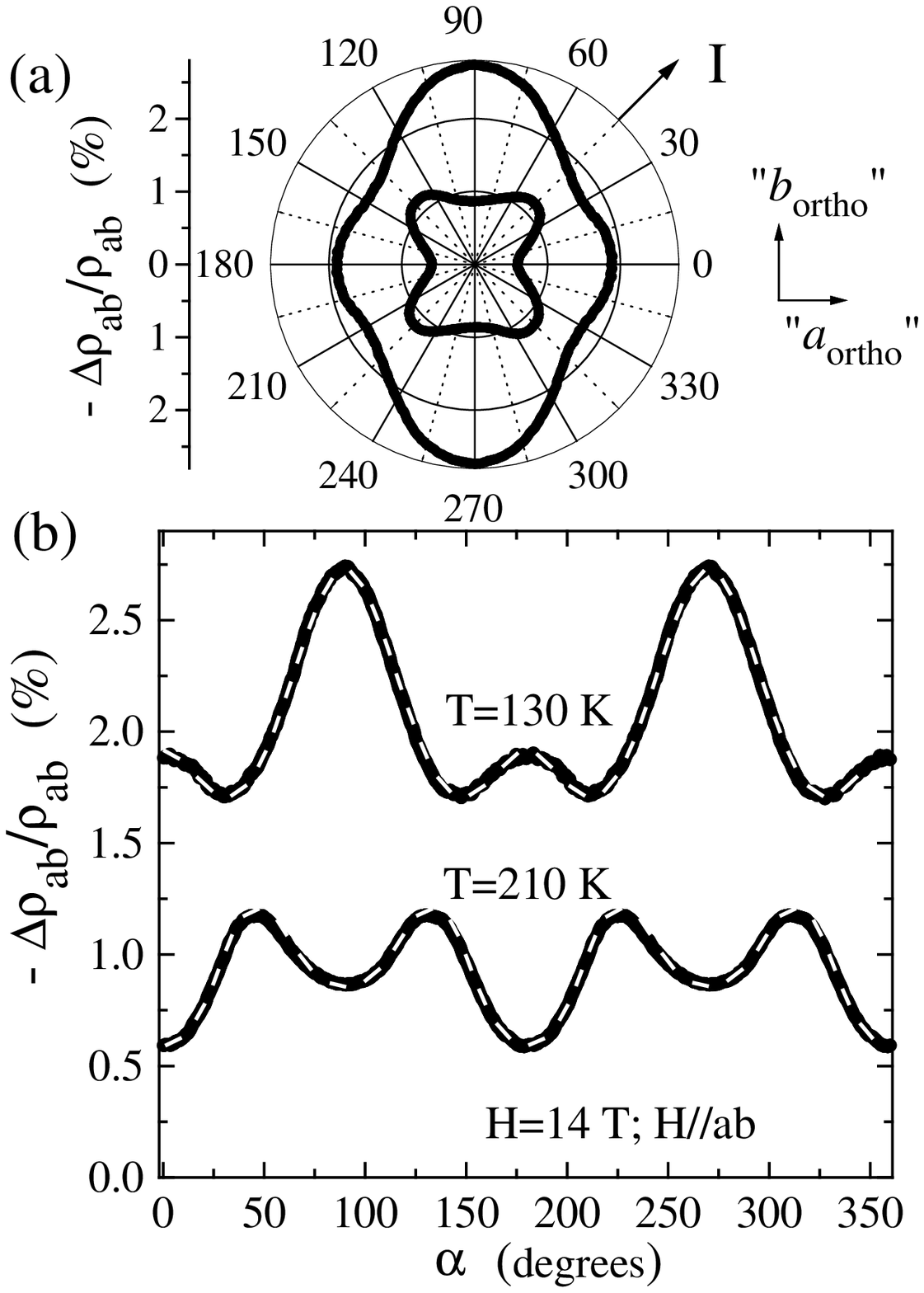}
\caption{Angular dependences of $\Delta\rho_{ab}/\rho_{ab}$ at $T=130$ K
and 210 K in (a) polar and (b) linear coordinates. The white dashed lines
in (b) show the fits described in the text.}
\label{fig4}
\end{figure}

\begin{figure*}[!tb]
\includegraphics*[width=15.5cm]{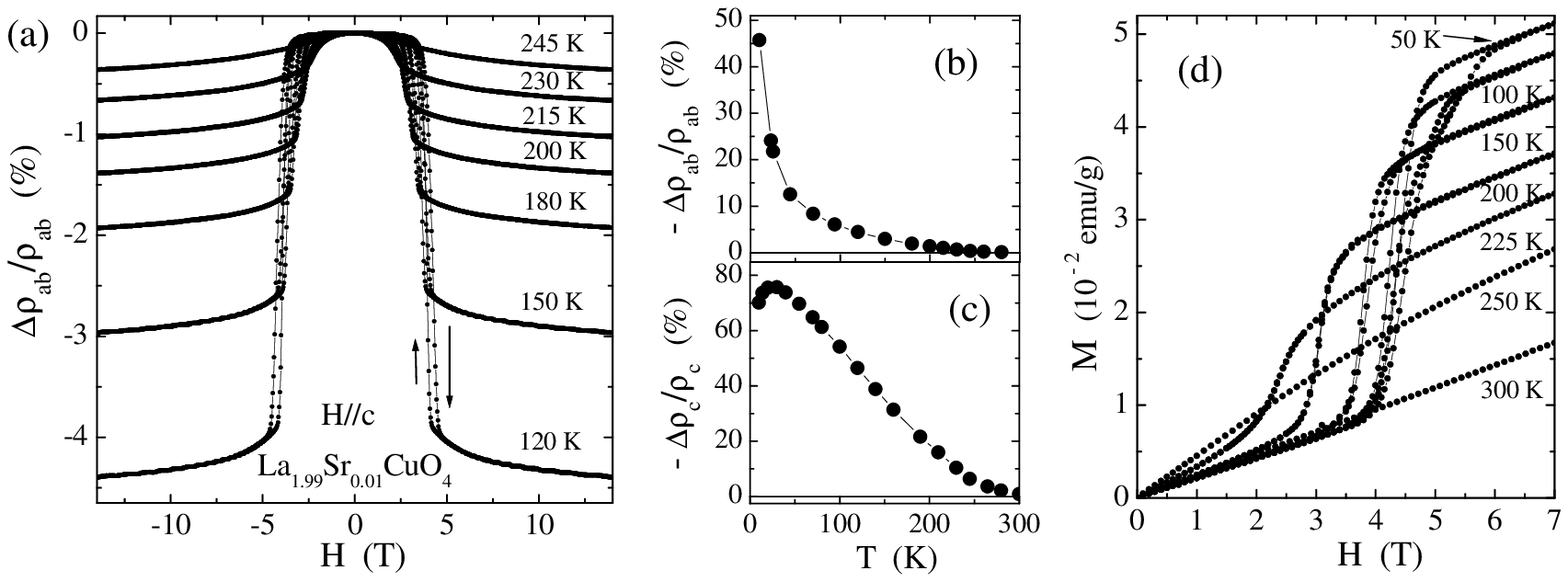}
\caption{(a) The MR in $\rho_{ab}$ of La$_{1.99}$Sr$_{0.01}$CuO$_4$ for
$H\parallel c$ at high temperatures; note the hysteresis marked by arrows.
(b, c) $T$ dependences of the MR at 14 T for $\rho_{ab}$ and $\rho_c$. (d)
Magnetization for $H\parallel c$ illustrating the WF transition.}
\label{fig5}
\end{figure*}

By now, the influence of magnetic fields on the AF spin order in undoped
La$_2$CuO$_4$ has been fairly well understood \cite{suscept,MR_WF}. At
zero field, spins are aligned almost perfectly along the $b$-axis, and
just slightly canted towards the $c$-axis, owing to the
Dzyaloshinskii-Moriya (DM) interaction; the weak FM moments induced by
this spin canting have opposite directions in adjacent CuO$_2$ planes so
that no net moment is observed at zero field. When high enough magnetic
fields are applied along the $c$-axis, the weak FM moments in every second
CuO$_2$ plane switch their orientation through a first-order transition
\cite{MR_WF,MR_flop}, which is manifested in a step-like increase in the
magnetization [Fig. 5(d)]. Since the direction of the canted moments is
uniquely linked with the local {\it phase} of the AF order, this {\it
phase} also switches in every second CuO$_2$ plane. On the other hand,
when $H \parallel b$ is applied, the weak FM moments, which are confined
to the $bc$ plane due to the DM vector ${\bf D}\parallel a$
\cite{suscept}, smoothly rotate from the $c$ to $b$ direction to become
parallel to the field. In any case, a magnetic field applied within the
$bc$ plane eventually aligns all the weak FM moments and unifies the phase
of the AF order over the crystal; note that since weak FM moments are
confined to the $bc$ plane, the field $H \parallel a$ can hardly alter the
spin order.

Apparently, the anisotropic magnetic-field effect on the spin order
correlates well with what we see in the field and angular dependences of
the MR. Therefore, it is natural to assert that the resistivity in LSCO is
somehow affected by the pattern of the phase of the AF order; in fact, a
large MR in La$_2$CuO$_{4+\delta}$ reported by Thio {\it et al.}
\cite{MR_WF,MR_flop} was attributed to an extraordinary sensitivity of the
{\it out-of-plane} conductivity to the relative spin ordering in adjacent
CuO$_2$ planes \cite{MR_flop,MRtheory} -- the behavior reminiscent of the
intrinsic spin-valve effects in manganites \cite{Tokura}. What is
important in the present data is, however, that the {\it in-plane}
resistivity shows clear and large changes upon unifying the phase of the
AF order. Given the quasi-2D conduction in our LSCO crystals
($\rho_c/\rho_{ab}\sim 10^3$), the {\it in-plane} transport can hardly be
sensitive to relative phases of the AF order in adjacent CuO$_2$ planes,
and therefore it must be the phase changes {\it within} the CuO$_2$ planes
and the removal thereof that is responsible for the peculiar MR in
$\rho_{ab}$. This means that the MR behavior in $\rho_{ab}$ gives evidence
that each CuO$_2$ plane intrinsically contains a set of anti-phase
boundaries in zero field.

Given that the anti-phase boundaries exist in the CuO$_2$ planes, an
important question is whether the holes are trapped in those boundaries.
Although theoretical calculations show they do \cite{theory2} (they even
suggest that it is the charges that dictate the anti-phase structure), it
is desirable to draw a conclusion from experiments. In this regard, the
temperature dependence of $\Delta\rho_{ab}/\rho_{ab}$ [Fig. 5(b)] is very
useful: $|\Delta\rho_{ab}/\rho_{ab}|$ is quite small when the conduction
is metal-like, but grows dramatically in the low-temperature insulating
region. If the holes are uniformly distributed and the anti-phase
boundaries are working as {\it scatterers} of holes, the removal of these
boundaries should primarily affect the scattering rate of holes; such a
change should have more significant effect in the metal-like regime rather
than in the insulating regime, where the conduction is governed not by the
scattering but by hopping. If, on the other hand, the primary function of
the boundaries is to confine holes, the effect of $H \parallel c$ is
expected to be more drastically observed in the insulating regime, since
the confinement potential changes. Clearly, our data indicate that the
latter is the case, which means that in zero-field the holes are confined
in the anti-phase boundaries, forming the charge stripes.

If one accepts the above conclusion, a puzzling aspect of our data is that
the wiping out of the anti-phase boundaries with magnetic fields results
in a noticeable {\it decrease} in $\rho_{ab}$; this appears to contradict
the usual notion that the anti-phase stripes are formed to facilitate the
hole motion in Mott insulators \cite{theory}. One possibility that
resolves this puzzle is that the stripes do not ``evaporate" in high
magnetic fields but change into in-phase stripes. In fact, theoretical
calculations show \cite{theory2} that the domain-wall formation is favored
over the uniform distribution of charges, and it appears \cite{theory3}
that the energetics to determine whether the domain walls (stripes) are
in-phase or anti-phase are rather subtle; thus, it is actually likely that
the in-phase stripes are formed when the anti-phase stripes are
prohibited. To understand the better conduction through the in-phase
stripes compared to the anti-phase stripes is probably more challenging,
but this might be resolved if one allows the hole filling in the stripes
to be different in the two states \cite{theory3}, which naturally changes
the conductivity through the stripes. While this last point is already
highly speculative, our experimental data offer a good testing ground for
theories of stripes in the cuprates.

\begin{acknowledgments}
We thank K. Segawa for invaluable technical assistance, and S. A. Kivelson
and I. Tsukada for fruitful discussions.
\end{acknowledgments}


\begin{thebibliography}{99}
\bibitem{note}
It is not surprising that the $H$-dependences of $\rho_{ab}$ and $\rho_c$
are different, since the transport mechanisms along and transverse to
CuO$_2$ planes are fundamentally different.
\end{thebibliography}
\end{document}